\documentclass[aps,prd,floatfix,groupedaddress,nofootinbib,superscriptaddress,twocolumn]{revtex4-1}

\usepackage{amsmath,amssymb,bm,color,float,graphicx,mathrsfs}
\usepackage[hidelinks=true]{hyperref}
\hypersetup{colorlinks=true,linkcolor=blue,citecolor=blue,urlcolor=blue}
\usepackage{Macro}

\def\T{\Theta}
\def\GW{{\rm GW}}

\begin{document}

\title{CMB Constraints on the Stochastic Gravitational-Wave Background at Mpc scales}

\author{Toshiya Namikawa}
\affiliation{Department of Applied Mathematics and Theoretical Physics, University of Cambridge, Wilberforce Road, Cambridge CB3 0WA, Unite Kingdom}
\author{Shohei Saga}
\affiliation{Center for Gravitational Physics, Yukawa Institute for Theoretical Physics, Kyoto University, Kyoto 606-8502, Japan}
\author{Daisuke Yamauchi}
\affiliation{Faculty of Engineering, Kanagawa University, Kanagawa 221-8686, Japan}
\author{Atsushi Taruya}
\affiliation{Center for Gravitational Physics, Yukawa Institute for Theoretical Physics, Kyoto University, Kyoto 606-8502, Japan}
\affiliation{Kavli Institute for the Physics and Mathematics of the Universe, The University of Tokyo Institutes for Advanced Study (UTIAS), The University of Tokyo, Chiba 277-8583, Japan}

\date{\today}

\begin{abstract}
We present robust constraints on the stochastic gravitational waves (GWs) at Mpc scales from the cosmic microwave background (CMB) data. CMB constraints on GWs are usually characterized as the tensor-to-scalar ratio, assuming specifically a power-law form of the primordial spectrum, and are obtained from the angular spectra of CMB. Here, we relax the assumption of the power-law form, and consider to what extent one can constrain a monochromatic GW at shorter wavelengths. Previously, such a constraint has been derived at the wavelengths larger than the resolution scale of the CMB measurements, typically above $100\,\Mpc$ (below $10^{-16}\,\Hz$ in frequency). However, GWs whose wavelength is much shorter than $100\,\Mpc$ can imprint a small but non-negligible signal on CMB anisotropies at observed angular scales, $\l<1000$. Here, using the CMB temperature, polarization, and lensing data set, we obtain the best constraints to date at $10^{-16}-10^{-14}\,\Hz$ of the GWs produced before the time of decoupling, which are tighter than those derived from the astrometric measurements and upper bounds on extra radiations. In the future, the constraints on GWs at Mpc scales will be further improved by several orders of magnitude with the precision $B$-mode measurement on large scales, $\l<100$. 
\end{abstract} 

\preprint{YITP-19-20}
\keywords{cosmology, cosmic microwave background, gravitational waves}

\maketitle


\section{Introduction} \label{sec:intro}

The stochastic gravitational wave (GW) background has been constrained by multiple observations. The cosmic microwave background (CMB) temperature and polarization observations have provided the tightest constraints on the GW background at very low frequencies, $f\alt 10^{-16}\,\Hz$ \cite{BKP,BKX,P18:main}. The upper bound on the amplitude of the primordial GW power spectrum has been translated to that on the stochastic GW background at scales larger than the angular resolution of CMB experiments, $k\alt 0.1\,\Mpc^{-1}$, which is equivalent to $f\alt 10^{-16}\,\Hz$ in frequency. Combining CMB and LSS data, the GW background is also constrained at higher frequencies, $f=10^{-16}-10^{-11}\,\Hz$, via the upper bound on the extra radiation at the time of CMB decoupling \cite{Smith:2006prl,Smith:2006prd,Pagano:2016}. Other astrophysical observations have put constraints on the GW background at roughly the same frequency range. References.~\cite{Gwinn:1996gv} and \cite{Darling:2018} use the motion (astrometry) of quasars and radio sources, respectively, to constrain the GW energy density. Reference~\cite{Titov:2011} derives the constraint on GWs from the secular aberration of the extragalatic radio sources caused by the rotation of the Solar System barycenter around the Galactic center. The stochastic GW background at higher frequencies, $f\agt 10^{-11}\,\Hz$, is also constrained by many other direct and indirect observations such as the Pulsar Timing Array \cite{Lasky:2016:PTA}, big bang nucleosynthesis (BBN) \cite{Henrot-Versille:2014jua}, and the Laser Interferometer Gravitational-Wave Observatory (LIGO) \cite{LIGO:2017:omegagw}. The absence of the primordial black hole also leads to the upper bound on the GW background at a broad range of frequencies \cite{Nakama:2016enz}. 

In this paper, we revisit the CMB constraints on the energy density of the stochastic GW background at $k\alt 10\,\Mpc^{-1}$ based on the upper bound on the amplitude of the primordial tensor power spectrum. The GW constraints by the primordial tensor power spectrum have been discussed only at the CMB scales, $k\alt 0.1\,\Mpc^{-1}$ (see e.g. \cite{LIGO:2017:omegagw}). This is because a finite angular resolution of CMB maps limits the observable range of the CMB angular multipole to $\l\alt 1000$, and the observable scale is restricted to $k\simeq \l/\chi_*\alt 0.1\,\Mpc^{-1}$ where $\chi_*\sim 10^4$ is the comoving distance to the CMB last scattering surface. The CMB spectrum is most sensitive to the primordial GWs at $k\alt 0.01$ Mpc$^{-1}$ \cite{Hiramatsu:2018nfa}. However, the CMB data can be used to constrain the stochastic GW background at small scales compared to the CMB scale, $k\agt 10$ Mpc$^{-1}$. The CMB anisotropies and lensing at large-angular scales come from the GW perturbations at low redshifts. Although such contributions are very small, we find that the upper limits on the large-scale CMB fluctuations provide tighter constraints on the GW energy density at $k=0.1-10$ Mpc$^{-1}$ than those from other existing constraints at the same scales. 

\begin{figure*}[tb]
\bc
\includegraphics[width=180mm,height=54mm,clip]{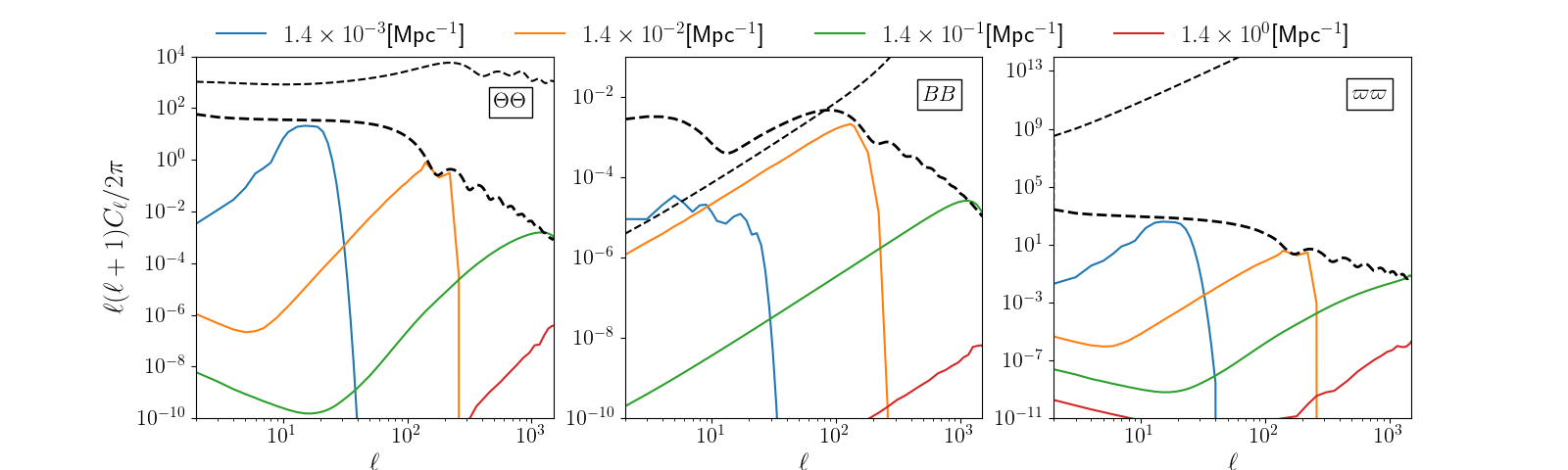}
\caption{
The angular spectra of the CMB temperature, $B$-mode and lensing curl-mode with varying the central frequency, $k_c$ (solid lines). The amplitude of the power spectra is given by $A_{\rm GW}=rA_s$ where the tensor-to-scalar ratio is chosen as the current best upper bound, $r=0.06$ \cite{BKX}, and the scalar amplitude at $k=0.05$ Mpc$^{-1}$ is consistent with the latest Planck cosmology, $A_s=2.1\times 10^{-9}$. The gray dashed lines in the temperature and $B$ mode spectra show the contributions from the scalar perturbations, while that in the curl-mode spectrum shows the reconstruction noise of the Planck observation. The black dashed lines show the inflationary GW contributions with $r=0.1$. 
}
\label{fig:aps}
\ec
\end{figure*}

To constrain the energy density of the stochastic GW background using CMB angular spectra, we need to assume a cosmological model for the scalar perturbations, though the degeneracy between the cosmological parameters and GW energy density would be small due to the difference of the angular scale dependence. In this respect, the GW constraints obtained by the CMB angular spectra depend on the model of the scalar perturbations. A less model-independent way to constrain the GW energy density by CMB observations is to use the curl-mode of the CMB deflection angle which has been discussed in several papers \cite{Cooray:2005hm,Sarkar:2008ii,Namikawa:2014:gwcurl,Saga:2015}. In the standard cosmology, the curl-mode is consistent with $0$ within the current measurement error of Planck. In this paper, we use the curl-mode to constrain the GW energy density as a more robust way than using the CMB angular spectra.

This paper is organized as follows. In Sec.~\ref{sec:aps}, we begin by discussing the CMB power spectra generated by small-scale GWs, and see their typical behaviors at large angular scales. Then, Section~\ref{sec:method} describes the data and our method to derive the constraints on small-scale GWs. Section~\ref{sec:results} presents our main results, i.e., the upper bound on the energy density of stochastic GWs, together with future forecast. Finally, Sec.~\ref{sec:summary} summarizes our work.

\begin{figure*}[tb]
\bc
\includegraphics[width=180mm,height=54mm,clip]{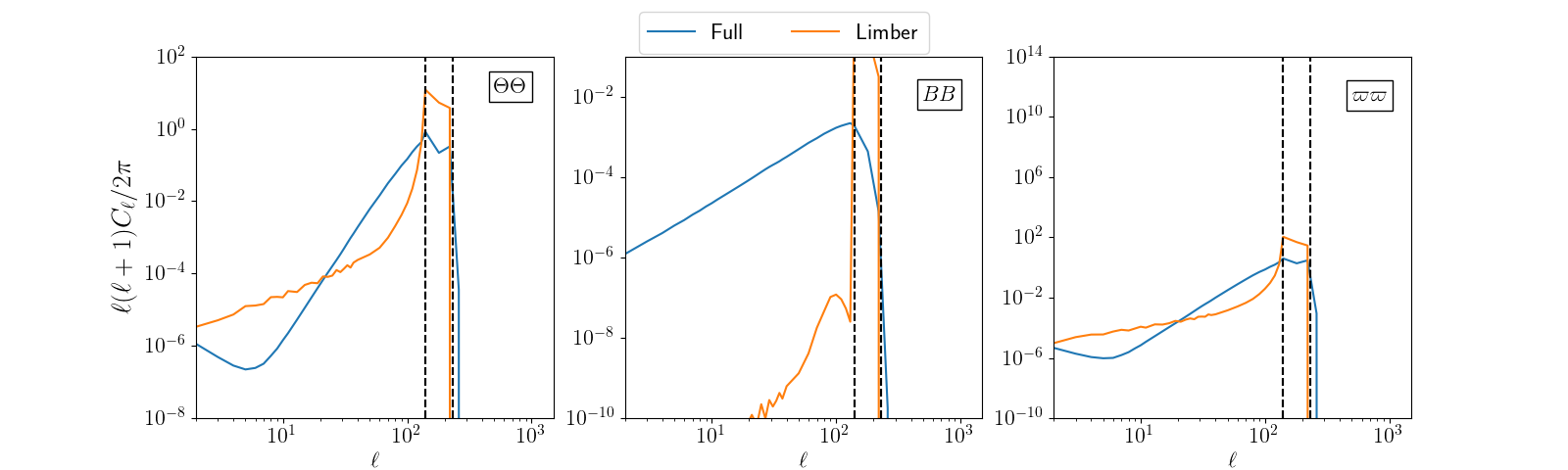}
\caption{
The angular spectra of the CMB temperature, $B$ mode and lensing curl-mode from a top-hat primordial GW power spectrum with $0.01\Mpc^{-1}\leq k\leq 0.018\Mpc^{-1}$ using the Limber approximation, $\l=k\chi$. The amplitude of the power spectra is the same as that in Fig.~\ref{fig:aps}. The vertical dashed lines show $\l=0.01\chi_*\,\Mpc^{-1}$ and $\l=0.018\chi_*\,\Mpc^{-1}$. For comparison, we also show the results without the Limber approximation (blue).
}
\label{fig:aps:limber}
\ec
\end{figure*}

\section{Angular spectrum} \label{sec:aps}

\begin{figure*}[tb]
\bc
\includegraphics[width=180mm,height=54mm,clip]{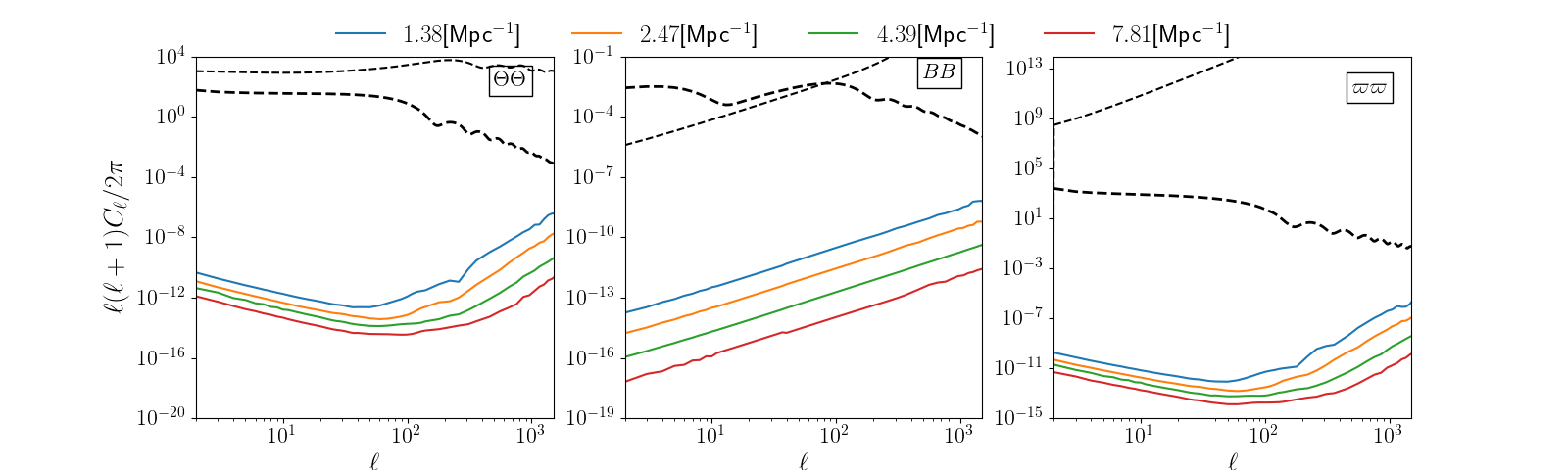}
\caption{
Same as Fig.~\ref{fig:aps} but for $k_c>1$ Mpc$^{-1}$. 
}
\label{fig:aps:highk}
\ec
\end{figure*}

CMB experiments observe the CMB temperature, $\T$, and Stokes Q/U maps at each pixel on the unit sphere. The Stokes Q/U maps are decomposed into the $E$/$B$ modes by the parity symmetry \cite{Kamionkowski:1996zd,Seljak:1996gy}. We then obtain the angular spectra of the temperature and $B$ modes by squaring the harmonic coefficients of the CMB anisotropies. 

The CMB primary anisotropies are distorted by gravitational lensing from the large-scale structure \cite{Lewis:2006fu,Hanson:2009kr}. The lensing distortion is described by a remapping of the CMB fluctuations at the CMB last scattering by a deflection angle, $\bm{d}$. The lensing effect leads to mode coupling between different CMB multipoles \cite{Hanson:2009gu}. This mode couplings can be used to reconstruct a map of the curl-mode of the deflection angle, $\curl=(\star\bn)\cdot\bm{d}$, from an observed CMB map (e.g., \cite{Okamoto:2002ik,Namikawa:2011:curlrec}), where $\star$ is the rotation operator for a two dimensional vector \cite{Namikawa:2011:curlrec}. We then measure the curl-mode spectrum. 

Given the initial dimensionless power spectrum of GWs, $\Delta_t(k)$, the CMB and curl-mode spectra are theoretically computed, with a help of the CMB Boltzmann code, as 
\al{
    C_\l^{XX} &= 4\pi \Int{}{\ln k}{} \Delta_t(k) \notag \\ 
    \times & \Int{}{\chi}{} j_\l(k\chi) S^X(k,\chi)\Int{}{\chi'}{} j_\l(k\chi')S^X(k,\chi') \, \label{Eq:Cl}
}
with $X=\Theta$, $B$ or $\curl$. In this paper, the source functions, $S^X(k,\chi)$, and angular spectra $C_\l^{XX}$ are computed, modifying CAMB \cite{Lewis:1999bs}. To obtain a generic constraint on stochastic GWs in a rather model-independent manner, we consider a monochromatic GW given at the wavenumber, $k_{\rm c}$. The dimensionless power spectrum of GW is then given by
\al{
	\Delta_t(k) = \begin{cases} A_{\rm GW}/(2\epsilon) & (|k/k_{\rm c}-1|\leq \epsilon) \\ 0 & (\text{otherwise}) \end{cases}
	\,.
}
Here, $A_{\rm GW}$ is the amplitude of GW, and $\epsilon$ characterizes the width of the GW spectrum in wavenumber. We divide the wavenumber between $10^{-4}$Mpc$^{-1}\leq k\leq 10$Mpc$^{-1}$ into logarithmically equal $20$ bins. We checked that our result remains unchanged even if we change the bin number to a larger value. 

Figure~\ref{fig:aps} shows the angular spectra of the CMB temperature (left), $B$-mode (middle), and curl-mode (right) for various wavenumber $k_{\rm c}$. The amplitude of the power spectra is given by $A_{\rm GW}=r\,A_s$, where the tensor-to-scalar ratio $r$ is set to be the current best upper bound, $r=0.06$ \cite{BKX}, with the scalar amplitude at $k=0.05$ Mpc$^{-1}$ determined by the latest Planck cosmology, $A_s=2.1\times 10^{-9}$ \cite{P18:main}. The large-scale Fourier modes ($k_c\ll 1$ Mpc$^{-1}$) contribute to the power spectrum at large scales. Similarly, the small-scale Fourier modes ($k_c\gg 1$ Mpc$^{-1}$) mostly generate the small-scale fluctuations. The large-scale fluctuations from such small-scale Fourier modes are typically small but their contribution is not exactly $0$. 

To elucidate the low-$\l$ behaviors shown in Fig.~\ref{fig:aps}, one  may compare the exact calculation with the Limber approximation. In the Limber approximation, the source function $S^X$ and power spectrum $\Delta_t$ in the integrand of \eq{Eq:Cl} are assumed to be a smooth function of $k$. Then, the integral convolving spherical Bessel function over $k$ exhibits a heavy cancellation, which results in a rather simplified form of the angular spectrum (see e.g. \cite{Loverde:2008,Lewis:2006fu}); 
\al{
    C_\l^{XX} &\simeq \frac{2\pi^2}{\l^3}\Int{}{\chi}{}\chi\Delta_t\Bigl(k=\frac{\l}{\chi}\Bigr)\,\left[S^X\Bigl(k=\frac{\l}{\chi},\chi\Bigr)\right]^2 
    \, \label{Eq:Cl-Limber}
}
which tells us that the amplitude of $C_\l^{XX}$ is determined by the contribution of GWs at the wavenumber $k=\ell/\chi$ projected along the line-of-sight (i.e., $\chi$-integral). Strictly speaking, the above equation is inadequate in our case because the integrand contains the top-hat primordial power spectrum. Further, the tensor transfer function has oscillatory behaviors, which, combining with spherical Bessel function, leads to a rather nontrivial cancellation. Nevertheless, Eq.~\eqref{Eq:Cl-Limber} can be used for a qualitative understanding of the angular spectrum.

In Fig.~\ref{fig:aps:limber}, we specifically set the top-hat GW spectrum to the one centered at $k_{\rm c}=0.014\,\Mpc^{-1}$ with the width $\Delta k=0.008\,\Mpc^{-1}$, and plot the angular spectra with and without the Limber approximation. Then, in all cases, the results with Limber approximation exhibit a sharply peaked structure around $\ell_*\equiv k_{\rm c}\chi_*$, indicated by the two vertical dashed lines, where $\chi_*$ is the comoving distance to the last scattering surface of CMB. At higher multipoles of $\l>\l_*$, the amplitudes rapidly falls off, consistent with exact calculations. These behaviors basically come from the nature of photon radiative transfer encapsulated in the function  $S^{\rm X}$. On the other hand, looking at the lower multipoles of $\l_*<\l$, while the Limber approximation predicts a rather suppressed $B$-mode spectrum that fails to reproduce the exact calculation, the predictions of the temperature and curl-mode spectra show a rather long tail, which qualitatively explains the behaviors in the exact calculations. Recall that in the Limber approximation, the top-hat GWs peaked at $k_{\rm c}$ can contribute to $C_\l$ only at the multipole $\l=k_{\rm c}\,\chi$, Fig.~\ref{fig:aps:limber} implies that the low-$\l$ behaviors of the exact calculation in the temperature and curl-mode spectrum mainly comes from the low-$z$ GW contributions (i.e., $\chi\lesssim\chi_*$), whereas a non-negligible amount of the high-$z$ GWs plays a role to determine the low-$\l$ amplitude of the $B$-mode spectrum.

Having confirmed the typical behaviors of the angular spectra, we further consider  the small-scale GWs, and plot in Fig.~\ref{fig:aps:highk} the angular spectra for $k_{\rm c}>1$ Mpc$^{-1}$. The results are compared with the contributions from the scalar perturbations or the reconstruction noise. As we have seen in Figs.~\ref{fig:aps} and \ref{fig:aps:limber}, no appreciable low-$\l$ tail is developed for the $B$-mode spectrum, since the polarization is only generated at the reionization and recombination. Figure~\ref{fig:aps:highk} suggests that the $B$-mode constraint on the small-scale GWs is basically limited by the angular resolution of CMB observations. On the other hand, the temperature and curl-mode power spectra exhibit a low-$\l$ tail that is more prominent and is rather enhanced compared to the results in Fig.~\ref{fig:aps}. This implies that large angle CMB data can still give a meaningful constraint on small-scale GWs. 

\section{Data and Method} \label{sec:method}

Here, we explain the data and the analysis method to constrain the stochastic GWs, particularly paying attention at small scales. In our analysis, we use the CMB temperature spectrum, $C_\l^{\T\T}$, measured by Planck \cite{P18:main}, the $B$-mode spectrum, $C_\l^{BB}$, by BICEP/Keck Array \cite{BKX}, and curl-mode spectrum, $C_L^{\curl\curl}$, by Planck \cite{P13:phi}. Note that the constraints from the temperature-$E$ cross and $E$-mode autospectra measured by Planck turn out to be statistically insignificant compared to that obtained by the temperature spectrum. In this paper, therefore, we only present the results from the temperature, $B$-mode and curl-mode spectra. We checked that our constraints remain unchanged even if we add other $B$-mode spectra measured by POLARBEAR \cite{PB17:BB} and SPTpol \cite{Keisler:2015hfa}. 

Provided the data, the constraint on the amplitude of stochastic GW, $A_{\rm GW}$, assuming the monochromatic wave with wavenumber $k_{\rm c}$, is obtained by minimizing the likelihood function $\mC{L}$. We adopt here the Gaussian likelihood function, 
\al{
	-2\ln \mC{L} (A_{\rm GW}) = \sum_{b=1}^n \frac{[\hC^{XY}_b-C^{XY,{\rm fid}}_b(A_{\rm GW})]^2}{(\sigma^{XY}_b)^2}
	\,, \label{Eq:likelihood}
}
where the subscript $b$ indicates the index of the multipole bins, and $n$ is the number of the multipole bins. The label $XY$ implies $\T\T$, $BB$ or $\curl\curl$. The power spectrum $\hC_b^{XY}$ is the measured binned spectrum, and $C_b^{XY,{\rm fid}}(A_{\rm GW})$ is the theoretical prediction having a specific GW amplitude $A_{\rm GW}$. Finally, $\sigma_b^{XY}$ is the measurement error of the angular spectrum provided by the CMB experiments described above. The multipole ranges used in the likelihood analysis are $2\leq\l\leq 2508$ for temperature, $37\leq\l\leq 332$ for $B$-mode, and $2\leq\l\leq2020$ for curl-mode spectra, respectively, at which the data are validated. 

The observed temperature spectrum has contributions from both the scalar and tensor perturbations. In the $B$-mode spectrum, the gravitational lensing effect generates the $B$ mode even if there is only the scalar perturbation \cite{Zaldarriaga:1998ar}. Thus, we simultaneously need to model or constrain the non-GW contributions to constrain GWs in the angular spectra. In our analysis, we simply subtract the contributions of the scalar perturbations from the measured spectrum, assuming the Planck best-fit $\Lambda$CDM model, since the degeneracy between the GW amplitude and cosmological parameters would be small and does not increase the upper bound by more than an order of magnitude. 

The upper bound on $A_{\rm GW}$ obtained from the above analysis is then translated to the GW fractional energy density defined as
\al{
	\Omega_\GW(k) &\equiv \frac{1}{\rho\rom{c}}\D{\rho_\GW}{\ln k}
		\biggl|_{\eta =\eta_0}
	= \frac{\Delta_t(k)}{12H_0^2}
		\left(\PD{T(k,\eta_0)}{\eta}\right)^2
	\notag \\
	&= \frac{(A_\GW/2\epsilon)}{12H_0^2}\left(\PD{T(k,\eta_0)}{\eta}\right)^2 
	\,,\label{eq:Omega_GW}
}
at $|k/k_{\rm c}-1|\leq \epsilon$ and $0$ otherwise, where $\rho\rom{c}$ is the critical density of the Universe, and $H_0$ is the Hubble parameter today. The time derivative of the GW transfer function is computed by CAMB assuming no neutrino anisotropic stress \cite{Lewis:1999bs}. 

\section{Results} \label{sec:results}

\begin{figure}[tb]
\bc
\includegraphics[width=90mm,height=63mm,clip]{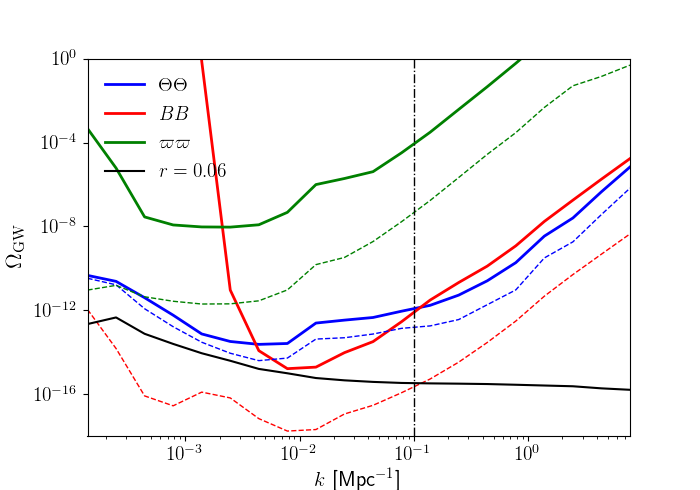}
\caption{
The current $95$\% upper bounds on the amplitude of the stochastic GW background at $k<10$ Mpc$^{-1}$ by the CMB observations. For reference, we show the energy density of the scale-invariant primordial GW with $r=0.06$ as a function of $k$ (black solid). The dashed lines are the expected upper bounds by a future CMB observation. The vertical dot-dashed line roughly corresponds to the maximum observable wavenumber determined by the resolution of CMB experiments. 
}
\label{fig:omegagw:cmb}
\ec
\end{figure}

Figure~\ref{fig:omegagw:cmb} shows the $95$\% C.L. upper bounds on the GW amplitudes at each scale using the current best measurements of the CMB angular spectra. At $k=0.01$ Mpc$^{-1}$, the best constraints come from the $B$-mode spectrum measured by BICEP2/Keck Array. At smaller scales, $k\agt0.1$ Mpc$^{-1}$, however, the temperature spectrum obtained by Planck provides the constraints on the GW amplitudes comparable to that from the $B$-mode spectrum. At large scales, the constraint by the $B$-mode spectrum becomes very weak because the $B$-mode spectrum is not measured at $\l\alt 30$. 

Figure~\ref{fig:omegagw:cmb} also plots the expected constraints on the GW spectrum amplitude by future CMB observations. In this forecast, we assume an idealistic future CMB experiment, i.e., a full sky observation, $1\mu$K-arcmin white noise with $1$ arcmin Gaussian beam. We also assume no foregrounds and $90$\% of the lensing $B$ mode is removed. Thus, the results provide the expected constraints in the most optimistic case. In the future, the constraints from the temperature spectrum are not significantly improved since the current temperature measurement is already dominated by the scalar perturbations at $\l<2000$. On the other hand, the $B$ mode and curl-mode are still dominated by the instrumental and reconstruction noise, respectively. The constraints are improved if the noise contribution is reduced in the future. The tightest constraints are obtained from the measurement of the $B$-mode spectrum.

\section{Summary and discussion} \label{sec:summary}

\begin{figure}[tb]
\bc
\includegraphics[width=90mm,height=63mm,clip]{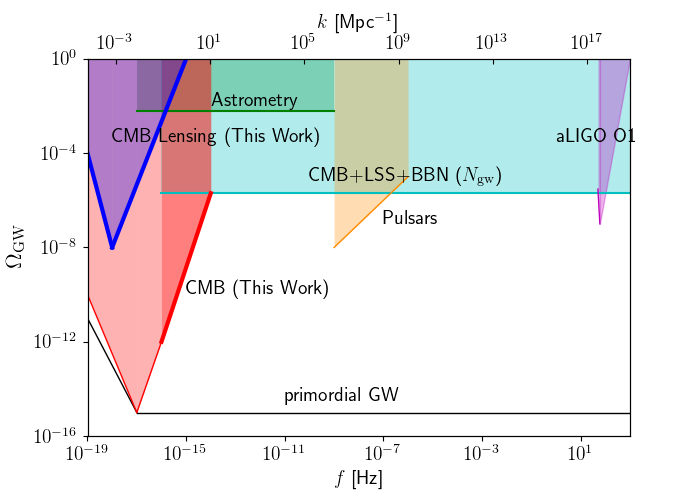}
\caption{
Summary of the current status of the upper bounds on the energy density of the stochastic GW background. The thick red and blue shaded regions correspond to the constraints obtained in this paper using the CMB angular spectra and curl-mode of the gravitational lensing, respectively. We also show other constraints from the astrometry of radio sources (Astrometry) \cite{Darling:2018}, extra radiation before CMB decoupling (CMB+LSS+BBN) \cite{Pagano:2016,Kohri:2018awv}, Pulsar Timing (Pulsars) \cite{Lasky:2016:PTA}, and GW interferometers (aLIGO O1)\cite{LIGO:2017:omegagw}. 
}
\label{fig:omegagw:all}
\ec
\end{figure}

In this paper, using the currently available CMB data, we presented robust constraints on stochastic GWs at Mpc scales, which are far below the resolution limit of CMB measurements. The key point is that even the short-wavelength GWs can still give an impact on the power spectrum of CMB anisotropies at large angular scales, to which we cannot apply the Limber approximation due to the heavily oscillatory behaviors of GWs. Using the available CMB data of temperature, $B$-mode polarization, and lensing from Planck and BICEP/Keck array, we find that currently both the temperature and $B$-mode data put the constraint almost at the same level at $k\gtrsim0.1$\,Mpc$^{-1}$, summarized as 
\al{
    \Omega_{\rm GW}\lesssim 10^{-12}\left(\frac{k}{0.1\,\mbox{Mpc}^{-1}}\right)^3 \,, \label{Eq:const}
}
which is compared with other constraints shown in Fig.~\ref{fig:omegagw:all}. Note that \eq{Eq:const} is, strictly speaking, valid at $k\lesssim10$\,Mpc$^{-1}$, at which we stop constraining GW due to the heavy oscillations of the spherical Bessel function and source function in the integrand of \eq{Eq:Cl}. However, it is expected from the behaviors seen in Fig.~\ref{fig:aps:highk} that the scaling given at \eq{Eq:const} still holds at $k\gtrsim10$\,Mpc$^{-1}$, thus giving a meaningful constraint complementary to those coming from astrometric observation and thermal history of the Universe. 

In the future, a tighter constraint will be obtained from the $B$-mode polarization, and the expected constraint will be improved by several orders of magnitude. Note cautiously that the constraints coming from the temperature and $B$-mode data are model dependent in the sense that the constraints are obtained from the data subtracting the primary (temperature) or secondary ($B$-mode) signal created by the scalar perturbations, for which we assume $\Lambda$CDM model. In this respect, the curl-mode data give complementary information to the temperature and B-mode polarization, although the constraining power is significantly degraded. 

Finally, the constraints obtained here are useful in narrowing the parameters of several models that can produce the GWs at Mpc scales. The early universe scenarios producing a blue-tilted stochastic GW are obviously interesting targets to constrain (e.g., \cite{Fujita:2018ehq}). Apart from these, the constraint on cosmic string models is also worth consideration. The stochastic GW spectrum produced by the cosmic strings depends significantly on the model parameters and assumptions as shown in Refs.~\cite{Sanidas:2012ee,Ringeval:2017eww,Blanco-Pillado:2017oxo}, and thus constraining the GWs over a very broad frequency range is important to pin down the cosmic string models. Another possible source of GWs at Mpc scales is the ultralight axion motivated by string theory. Reference ~\cite{Kitajima:2018zco} discusses the possibility that ultralight axion produces a sizable amount of GWs through the nonlinear scalar field interaction induced by parametric resonance, and the produced GWs can have a peak with broad distribution at Mpc scales if the axion mass is around $m\sim10^{-22}$\,eV. This mass scale lies at the scales for which the axion can put an interesting imprint on the small-scale structure \cite{Hu:2000:axion,Marsh:2013,Schive:2014dra,Schwabe:2016rze,Hui:2016:axion}. Several works also discuss generation of GWs at Mpc scales assuming that the primordial GW power spectrum has a sharp peak \cite{Saito:2008:pbh,Kohri:2018awv}. The constraint on stochastic GWs obtained in this paper certainly helps to exclude a part of parameter regions that can produce a large GW amplitude, although a detailed analysis is still necessary to get a constraint, using properly the shape of the predicted GW spectrum.



\begin{acknowledgments}
We thank Kazuyuki Akitsu, Kazunori Kohri, Toshiki Kurita, Blake Sherwin and Alexander van Engelen for helpful comments. T. N. acknowledges the support from the Ministry of Science and Technology (MOST), Taiwan, R.O.C. through the MOST research project grants (Grant No. 107-2112-M-002-002-MY3). This research used resources of the National Energy Research Scientific Computing Center, which is supported by the Office of Science of the U.S. Department of Energy under Contract No. DE-AC02-05CH11231. This work was supported in part by MEXT/JSPS KAKENHI Grants No. JP15H05899, No. JP17H06359 and No. JP16H03977 (A. T.), Grant No. 17K14304 (D. Y.), and Grant-in-Aid for JSPS fellows Grant No. 17J10553 (S. S.).
\end{acknowledgments}

\appendix

\bibliographystyle{mybst}
\bibliography{main,exp}

\providecommand{\href}[2]{#2}\begingroup\raggedright\begin{thebibliography}{10}

\bibitem{BKP}
{\textsc{Bicep2} and {\it Planck} Collaborations}, {\it ``Joint Analysis of
  BICEP2/Keck Array and Planck Data''},  {\em \prl} {\bf 114} (2015) 101301,
  [\href{http://arxiv.org/abs/1502.00612}{{\tt arXiv:1502.00612}}].

\bibitem{BKX}
{\textsc{Bicep2} / {\it Keck Array} Collaboration}, {\it ``{\sc BICEP2} / {\it
  Keck Array} X: Constraints on Primoridal Gravitational Waves Using Planck,
  WMAP, and New BICEP2/Keck Observations through the 2015 Season''},  {\em
  \prl} {\bf 121} (2018) 221301, [\href{http://arxiv.org/abs/1810.05216}{{\tt
  arXiv:1810.05216}}].

\bibitem{P18:main}
{\textit{Planck} Collaboration}, {\it ``Planck 2018 results. VI. Cosmological
  parameters''},  {\em \aap} (2018)
  [\href{http://arxiv.org/abs/1807.06209}{{\tt arXiv:1807.06209}}].

\bibitem{Smith:2006prl}
T.~L. Smith, E.~Pierpaoli, and M.~Kamionkowski, {\it ``A new cosmic microwave
  background constraint to primordial gravitational waves''},  {\em \prl} {\bf
  97} (2006), no.~021301 [\href{http://arxiv.org/abs/astro-ph/0603144}{{\tt
  astro-ph/0603144}}].

\bibitem{Smith:2006prd}
T.~L. Smith, M.~Kamionkowski, and A.~Cooray, {\it ``Direct detection of the
  inflationary gravitational wave background''},  {\em \prd} {\bf 73} (2006),
  no.~023504 [\href{http://arxiv.org/abs/astro-ph/0506422}{{\tt
  astro-ph/0506422}}].

\bibitem{Pagano:2016}
L.~Pagano, L.~Salvati, and A.~Melchiorri, {\it ``New constraints on primordial
  gravitational waves from Planck 2015''},  {\em \plb} {\bf 760} (2016) 823,
  [\href{http://arxiv.org/abs/1508.02393}{{\tt arXiv:1508.02393}}].

\bibitem{Gwinn:1996gv}
C.~R. Gwinn, T.~M. Eubanks, T.~Pyne, M.~Birkinshaw, and D.~N. Matsakis, {\it
  ``{Quasar proper motions and low frequency gravitational waves}''},  {\em
  \apj} {\bf 485} (1997) 87--91,
  [\href{http://arxiv.org/abs/astro-ph/9610086}{{\tt astro-ph/9610086}}].

\bibitem{Darling:2018}
J.~Darling, A.~E. Truebenbach, and J.~Paine, {\it ``Astrometric Limits on the
  Stochastic Gravitational Wave Background''},  {\em \apj} {\bf 861} (2018),
  no.~2 113, [\href{http://arxiv.org/abs/1804.06986}{{\tt arXiv:1804.06986}}].

\bibitem{Titov:2011}
O.~Titov, S.~B. Lambert, and A.~M. Gontier, {\it ``{VLBI measurement of the
  secular aberration drift}''},  {\em \aap} {\bf 529} (2011) A91,
  [\href{http://arxiv.org/abs/1009.3698}{{\tt arXiv:1009.3698}}].

\bibitem{Lasky:2016:PTA}
P.~D. Lasky, C.~M.~F. Mingarelli, T.~L. Smith, J.~T. Giblin, E.~Thrane, D.~J.
  Reardon, R.~Caldwell, M.~Bailes, N.~D.~R. Bhat, S.~Burke-Spolaor, S.~Dai,
  J.~Dempsey, G.~Hobbs, M.~Kerr, Y.~Levin, R.~N. Manchester, S.~Oslowski,
  V.~Ravi, P.~A. Rosado, R.~M. Shannon, R.~Spiewak, W.~van Straten, L.~Toomey,
  J.~Wang, L.~Wen, X.~You, and X.~Zhu, {\it ``Gravitational-Wave Cosmology
  across 29 Decades in Frequency''},  {\em Phys. Rev. X} {\bf 6} (2016) 011035,
  [\href{http://arxiv.org/abs/1511.05994}{{\tt arXiv:1511.05994}}].

\bibitem{Henrot-Versille:2014jua}
S.~Henrot-Versille {\em et~al.}, {\it ``{Improved constraint on the primordial
  gravitational-wave density using recent cosmological data and its impact on
  cosmic string models}''},  {\em Class. Quant. Grav.} {\bf 32} (2015), no.~4
  045003, [\href{http://arxiv.org/abs/1408.5299}{{\tt arXiv:1408.5299}}].

\bibitem{LIGO:2017:omegagw}
{LIGO Scientific and Virgo Collaborations}, {\it ``Upper Limits on the
  Stochastic Gravitational-Wave Background from Advanced LIGO’s First
  Observing Run''},  {\em \prl} {\bf 118} (2017) 121101,
  [\href{http://arxiv.org/abs/1612.02029}{{\tt arXiv:1612.02029}}].

\bibitem{Nakama:2016enz}
T.~Nakama and T.~Suyama, {\it ``{Primordial black holes as a novel probe of
  primordial gravitational waves. II: Detailed analysis}''},  {\em \prd} {\bf
  94} (2016), no.~4 043507, [\href{http://arxiv.org/abs/1605.04482}{{\tt
  arXiv:1605.04482}}].

\bibitem{Hiramatsu:2018nfa}
T.~Hiramatsu, E.~Komatsu, M.~Hazumi, and M.~Sasaki, {\it ``{Reconstruction of
  primordial tensor power spectra from B-mode polarization of the cosmic
  microwave background}''},  {\em \prd} {\bf 97} (2018), no.~12 123511,
  [\href{http://arxiv.org/abs/1803.00176}{{\tt arXiv:1803.00176}}].

\bibitem{Cooray:2005hm}
A.~Cooray, M.~Kamionkowski, and R.~R. Caldwell, {\it ``Cosmic shear of the
  microwave background: The curl diagnostic''},  {\em \prd} {\bf 71} (2005)
  123527, [\href{http://arxiv.org/abs/astro-ph/0503002}{{\tt
  astro-ph/0503002}}].

\bibitem{Sarkar:2008ii}
D.~Sarkar, P.~Serra, A.~Cooray, K.~Ichiki, and D.~Baumann, {\it ``{Cosmic shear
  from scalar-induced gravitational waves}''},  {\em \prd} {\bf 77} (2008)
  103515, [\href{http://arxiv.org/abs/0803.1490}{{\tt arXiv:0803.1490}}].

\bibitem{Namikawa:2014:gwcurl}
T.~Namikawa, D.~Yamauchi, and A.~Taruya, {\it ``Future detectability of
  gravitational-wave induced lensing from high-sensitivity CMB experiments''},
  {\em \prd} {\bf 91} (2015), no.~4 043531,
  [\href{http://arxiv.org/abs/1411.7427}{{\tt arXiv:1411.7427}}].

\bibitem{Saga:2015}
S.~Saga, D.~Yamauchi, and K.~Ichiki, {\it ``Weak lensing induced by
  second-order vector mode''},  {\em \prd} {\bf 92} (2015) 063533,
  [\href{http://arxiv.org/abs/1505.02774}{{\tt arXiv:1505.02774}}].

\bibitem{Kamionkowski:1996zd}
M.~Kamionkowski, A.~Kosowsky, and A.~Stebbins, {\it ``A Probe of primordial
  gravity waves and vorticity''},  {\em \prl} {\bf 78} (1997) 2058--2061,
  [\href{http://arxiv.org/abs/astro-ph/9609132}{{\tt astro-ph/9609132}}].

\bibitem{Seljak:1996gy}
U.~Seljak and M.~Zaldarriaga, {\it ``Signature of gravity waves in polarization
  of the microwave background''},  {\em \prl} {\bf 78} (1997) 2054--2057,
  [\href{http://arxiv.org/abs/astro-ph/9609169}{{\tt astro-ph/9609169}}].

\bibitem{Lewis:2006fu}
A.~Lewis and A.~Challinor, {\it ``Weak gravitational lensing of the CMB''},
  {\em Phys. Rep.} {\bf 429} (2006) 1--65,
  [\href{http://arxiv.org/abs/astro-ph/0601594}{{\tt astro-ph/0601594}}].

\bibitem{Hanson:2009kr}
D.~Hanson, A.~Challinor, and A.~Lewis, {\it ``Weak lensing of the CMB''},  {\em
  Gen. Rel. Grav.} {\bf 42} (2010) 2197--2218,
  [\href{http://arxiv.org/abs/0911.0612}{{\tt arXiv:0911.0612}}].

\bibitem{Hanson:2009gu}
D.~Hanson and A.~Lewis, {\it ``Estimators for CMB Statistical Anisotropy''},
  {\em \prd} {\bf 80} (2009) 063004,
  [\href{http://arxiv.org/abs/0908.0963}{{\tt arXiv:0908.0963}}].

\bibitem{Okamoto:2002ik}
T.~Okamoto and W.~Hu, {\it ``The angular trispectra of CMB temperature and
  polarization''},  {\em \prd} {\bf 66} (2002) 063008,
  [\href{http://arxiv.org/abs/astro-ph/0206155}{{\tt astro-ph/0206155}}].

\bibitem{Namikawa:2011:curlrec}
T.~Namikawa, D.~Yamauchi, and A.~Taruya, {\it ``Full-sky lensing reconstruction
  of gradient and curl modes from CMB maps''},  {\em \jcap} {\bf 1201} (2012)
  007, [\href{http://arxiv.org/abs/1110.1718}{{\tt arXiv:1110.1718}}].

\bibitem{Lewis:1999bs}
A.~Lewis, A.~Challinor, and A.~Lasenby, {\it ``Efficient Computation of CMB
  anisotropies in closed FRW models''},  {\em \apj} {\bf 538} (2000) 473--476,
  [\href{http://arxiv.org/abs/astro-ph/9911177}{{\tt astro-ph/9911177}}].

\bibitem{Loverde:2008}
M.~{Loverde} and N.~{Afshordi}, {\it ``{Extended Limber approximation}''},
  {\em \prd} {\bf 78} (Dec, 2008) 123506,
  [\href{http://arxiv.org/abs/0809.5112}{{\tt arXiv:0809.5112}}].

\bibitem{P13:phi}
{\textit{Planck} Collaboration}, {\it ``Planck 2013 results. XVII.
  Gravitational lensing by large-scale structure''},  {\em \aap} {\bf 571}
  (2014) A17, [\href{http://arxiv.org/abs/1303.5077}{{\tt arXiv:1303.5077}}].

\bibitem{PB17:BB}
{\textsc{Polarbear} Collaboration}, P.~A.~R. Ade, M.~Aguilar, Y.~Akiba,
  K.~Arnold, C.~Baccigalupi, D.~{Barron}, D.~{Beck}, F.~{Bianchini},
  D.~{Boettger}, J.~{Borrill}, S.~{Chapman}, Y.~{Chinone}, K.~{Crowley},
  A.~{Cukierman}, R.~{D{\"u}nner}, M.~{Dobbs}, A.~{Ducout}, T.~{Elleflot},
  J.~{Errard}, G.~{Fabbian}, S.~M. {Feeney}, C.~{Feng}, T.~{Fujino},
  N.~{Galitzki}, A.~{Gilbert}, N.~{Goeckner-Wald}, J.~C. {Groh}, G.~{Hall},
  N.~{Halverson}, T.~{Hamada}, M.~{Hasegawa}, M.~{Hazumi}, C.~A. {Hill},
  L.~{Howe}, Y.~{Inoue}, G.~{Jaehnig}, A.~H. {Jaffe}, O.~{Jeong}, D.~{Kaneko},
  N.~{Katayama}, B.~{Keating}, R.~{Keskitalo}, T.~{Kisner},
  N.~{Krachmalnicoff}, A.~{Kusaka}, M.~{Le Jeune}, A.~T. {Lee}, E.~M. {Leitch},
  D.~{Leon}, E.~{Linder}, L.~{Lowry}, F.~{Matsuda}, T.~{Matsumura},
  Y.~{Minami}, J.~{Montgomery}, M.~{Navaroli}, H.~{Nishino}, H.~{Paar},
  J.~{Peloton}, A.~T.~P. {Pham}, D.~{Poletti}, G.~{Puglisi}, C.~L. {Reichardt},
  P.~L. {Richards}, C.~{Ross}, Y.~{Segawa}, B.~D. {Sherwin}, M.~{Silva-Feaver},
  P.~{Siritanasak}, N.~{Stebor}, R.~{Stompor}, A.~{Suzuki}, O.~{Tajima},
  S.~{Takakura}, S.~{Takatori}, D.~{Tanabe}, G.~P. {Teply}, T.~{Tomaru},
  C.~{Tucker}, N.~{Whitehorn}, and A.~{Zahn}, {\it ``{A Measurement of the
  Cosmic Microwave Background $B$-Mode Polarization Power Spectrum at
  Sub-Degree Scales from 2 years of POLARBEAR Data}''},  {\em \apj} {\bf 848}
  (2017) 121, [\href{http://arxiv.org/abs/1705.02907}{{\tt arXiv:1705.02907}}].

\bibitem{Keisler:2015hfa}
{SPTpol Collaboration (Keisler, R. et al.)}, {\it ``Measurements of Sub-degree
  B-mode Polarization in the Cosmic Microwave Background from 100 Square
  Degrees of SPTpol Data''},  {\em \apj} {\bf 807} (2015) 151,
  [\href{http://arxiv.org/abs/1503.02315}{{\tt arXiv:1503.02315}}].

\bibitem{Zaldarriaga:1998ar}
M.~Zaldarriaga and U.~Seljak, {\it ``Gravitational lensing effect on cosmic
  microwave background polarization''},  {\em \prd} {\bf 58} (1998) 023003,
  [\href{http://arxiv.org/abs/astro-ph/9803150}{{\tt astro-ph/9803150}}].

\bibitem{Kohri:2018awv}
K.~Kohri and T.~Terada, {\it ``{Semianalytic calculation of gravitational wave
  spectrum nonlinearly induced from primordial curvature perturbations}''},
  {\em \prd} {\bf 97} (2018), no.~12 123532,
  [\href{http://arxiv.org/abs/1804.08577}{{\tt arXiv:1804.08577}}].

\bibitem{Fujita:2018ehq}
T.~Fujita, S.~Kuroyanagi, S.~Mizuno, and S.~Mukohyama, {\it ``{Blue-tilted
  Primordial Gravitational Waves from Massive Gravity}''},  {\em Phys. Lett.}
  {\bf B789} (2019) 215--219, [\href{http://arxiv.org/abs/1808.02381}{{\tt
  arXiv:1808.02381}}].

\bibitem{Sanidas:2012ee}
S.~A. Sanidas, R.~A. Battye, and B.~W. Stappers, {\it ``Constraints on cosmic
  string tension imposed by the limit on the stochastic gravitational wave
  background from the European Pulsar Timing Array''},  {\em \prd} {\bf 85}
  (2012) 122003, [\href{http://arxiv.org/abs/1201.2419}{{\tt
  arXiv:1201.2419}}].

\bibitem{Ringeval:2017eww}
C.~Ringeval and T.~Suyama, {\it ``{Stochastic gravitational waves from cosmic
  string loops in scaling}''},  {\em JCAP} {\bf 1712} (2017), no.~12 027,
  [\href{http://arxiv.org/abs/1709.03845}{{\tt arXiv:1709.03845}}].

\bibitem{Blanco-Pillado:2017oxo}
J.~J. Blanco-Pillado and K.~D. Olum, {\it ``{Stochastic gravitational wave
  background from smoothed cosmic string loops}''},  {\em \prd} {\bf 96}
  (2017), no.~10 104046, [\href{http://arxiv.org/abs/1709.02693}{{\tt
  arXiv:1709.02693}}].

\bibitem{Kitajima:2018zco}
N.~Kitajima, J.~Soda, and Y.~Urakawa, {\it ``{Gravitational wave forest from
  string axiverse}''},  {\em JCAP} {\bf 1810} (2018), no.~10 008,
  [\href{http://arxiv.org/abs/1807.07037}{{\tt arXiv:1807.07037}}].

\bibitem{Hu:2000:axion}
W.~Hu, R.~Barkana, and A.~Gruzinov, {\it ``Cold and fuzzy dark matter''},  {\em
  \prl} {\bf 85} (2000) 1158--1161,
  [\href{http://arxiv.org/abs/astro-ph/0003365}{{\tt astro-ph/0003365}}].

\bibitem{Marsh:2013}
D.~J.~E. Marsh and J.~Silk, {\it ``A Model For Halo Formation With Axion Mixed
  Dark Matter''},  {\em \mnras} {\bf 437} (2014) 2652--2663,
  [\href{http://arxiv.org/abs/1307.1705}{{\tt arXiv:1307.1705}}].

\bibitem{Schive:2014dra}
H.-Y. Schive, T.~Chiueh, and T.~Broadhurst, {\it ``{Cosmic Structure as the
  Quantum Interference of a Coherent Dark Wave}''},  {\em Nature Phys.} {\bf
  10} (2014) 496--499, [\href{http://arxiv.org/abs/1406.6586}{{\tt
  arXiv:1406.6586}}].

\bibitem{Schwabe:2016rze}
B.~Schwabe, J.~C. Niemeyer, and J.~F. Engels, {\it ``{Simulations of solitonic
  core mergers in ultralight axion dark matter cosmologies}''},  {\em \prd}
  {\bf 94} (2016), no.~4 043513, [\href{http://arxiv.org/abs/1606.05151}{{\tt
  arXiv:1606.05151}}].

\bibitem{Hui:2016:axion}
L.~Hui, J.~P. Ostriker, S.~Tremaine, and E.~Witten, {\it ``{Ultralight scalars
  as cosmological dark matter}''},  {\em \prd} {\bf 95} (2017) 043541,
  [\href{http://arxiv.org/abs/1610.08297}{{\tt arXiv:1610.08297}}].

\bibitem{Saito:2008:pbh}
R.~Saito and J.~Yokoyama, {\it ``Gravitational wave background as a probe of
  the primordial black hole abundance''},  {\em \prl} {\bf 102} (2009) 161101,
  [\href{http://arxiv.org/abs/0812.4339}{{\tt arXiv:0812.4339}}].

\end{thebibliography}\endgroup

\end{document}